\documentclass[10pt,twocolumn, conference]{IEEEtran}

\usepackage{amsmath}
\usepackage{amssymb}
\usepackage[dvips]{graphicx}
\usepackage{setspace}
\usepackage{epsfig}
\usepackage{amsmath}
\usepackage{amssymb}
\usepackage[dvips]{graphicx}
\usepackage{epsfig}
\usepackage{algorithm,algpseudocode,caption}

\newcommand{\tSNR}{\text{SNR}}

\newtheorem{theo}{Theorem}

\newcommand{\figsize}{0.45}

\newcommand{\sr}{\text{SR}}

\begin{document}
\IEEEoverridecommandlockouts
\title{Security in Cognitive Radio Networks}
\author{\authorblockN{Sami Akin}
\authorblockA{Institute of Communications Technology\\
Leibniz Universit\"{a}t Hannover\\
Hannover, Germany 30167\\
Email: sami.akin@ikt.uni-hannover.de}\thanks{This work was supported by the European Research Council under Starting Grant 306644.}}	
\date{}

\maketitle

\begin{abstract}
In this paper, we investigate the information-theoretic security by modeling a cognitive radio wiretap channel under quality-of-service (QoS) constraints and interference power limitations inflicted on primary users (PUs). We initially define four different transmission scenarios regarding channel sensing results and their correctness. We provide effective secure transmission rates at which a secondary eavesdropper is refrained from listening to a secondary transmitter (ST). Then, we construct a channel state transition diagram that characterizes this channel model. We obtain the effective secure capacity which describes the maximum constant buffer arrival rate under given QoS constraints. We find out the optimal transmission power policies that maximize the effective secure capacity, and then, we propose an algorithm that, in general, converges quickly to these optimal policy values. Finally, we show the performance levels and gains obtained under different channel conditions and scenarios. And, we emphasize, in particular, the significant effect of hidden-terminal problem on information-theoretic security in cognitive radios.
\end{abstract}

\section{Introduction}
Due to its spreading nature, security in wireless communications has become an important scrutiny for a long time, and it has been analyzed from several research perspectives including Information Theory \cite{ekrem}. In the earliest information-theoretic study conducted by Shannon \cite{shannon}, the wiretap channel in which a transmitter communicates with a legitimate receiver with the presence of an eavesdropper was characterized \cite{ekrem}. Considering the wiretap channel, Wyner \cite{wyner} showed that secure transmission to a legitimate destination can be achieved, and he defined secrecy capacity as the highest reliable rate from the transmitter to the legitimate receiver while keeping the eavesdropper completely puzzled about the transmitted signal. Following these studies, there have been recently a large number of investigations addressing information-theoretic security under various conditions in different channel models \cite{liang}-\cite{tekin}. For instance, secure communications and secrecy capacity in fading channels have been studied in \cite{liang} and \cite{gopala}.


Meanwhile, on the grounds of randomness and fluctuations in wireless environment, not only security but also quality-of-service (QoS) regarding delay and buffer constraints has been considered as a vital performance metric. Accordingly, effective capacity \cite{wu_negi} as a dual of effective bandwidth \cite{chang} has attracted a remarkable focus as it has identified the maximum constant arrival rate that a given time-varying service process can support while meeting QoS requirements. Subsequently, the authors in \cite{liu_chamberland}-\cite{sonia} meticulously researched effective capacity in several contexts. For example, it was investigated in cognitive radios under interference power limitations while regarding channel sensing results \cite{sami_1} and in cognitive radio relay channels under average and peak power constraints \cite{sonia}.

During the time the ever-increasing demand for higher data rates with stringent QoS requirements is inevitably leading to highly intense spectrum occupancy, cognitive radio has emerged as a solution to ameliorate spectrum inefficiency, and to provide better QoS. Since then, several complex transmission scenarios have been proposed and analyzed, i.e., cooperative strategies for cognitive radio networks attracted significant attention \cite{cooperative}. With the growing multiplicity, security challenges have become more sophisticated and multi-directional in different aspects \cite{cognitive_challenge}. Specifically, as a substance of information-theoretic security in cognitive radios, the outage probability of secrecy capacity of a primary user (PU) from a theoretical point of view has been studied \cite{shu}, and the relationship between multiple-input multiple-output channel secrecy rate and the cognitive radios under interference power constraints has been examined \cite{lan}.

In this paper, different than the aforementioned studies, we investigate the information-theoretic security of secondary users (SUs) in cognitive radio wiretap channels where SUs detect the activities of PUs imperfectly, and then perform data transmission under QoS and interference power constraints. The organization of the rest of the paper is as follows. In Section \ref{channel_model_and_power_constraints}, we describe the cognitive radio wiretap channel model, and define the interference power constraints considering channel sensing errors. In Section \ref{secrecy_rates_section}, we provide the instantaneous secrecy capacities that depend on channel sensing results and channel fading parameters. Furthermore, we identify channel states, and establish the effective secure capacity in Section \ref{effective_secure_rate_section}, and obtain the optimal transmission power policies that maximize this capacity in Section \ref{optimal_transmission_power_section}. Finally, we present the numerical results in Section \ref{numerical_results_section}, and we summarize our achievements in Section \ref{conclusion_section}.

\section{Channel Model and Power Constraints}\label{channel_model_and_power_constraints}
\subsection{Channel Model}
\begin{figure}
\begin{center}
\includegraphics[width=0.4\textwidth]{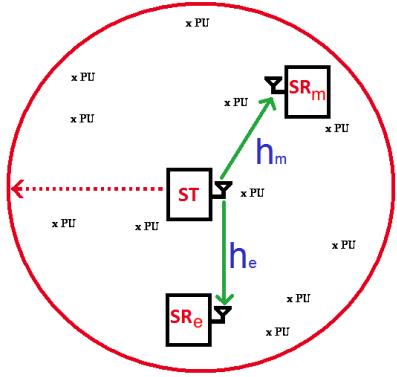}
\caption{Cognitive radio wiretap channel model.}\label{res:res_1}
\end{center}
\end{figure}

As shown in Figure \ref{res:res_1}, we consider a setting in which a single secondary transmitter, denoted by ST, performs communications with two secondary receivers (SRs) in the presence of possibly multiple PUs. We assume that ST sends confidential messages to one of the SRs, which is denoted by $\sr_{m}$. In this perspective, the other SR, which is denoted by $\sr_{e}$, can be considered as an eavesdropper. Nevertheless, it is not necessarily a malicious one in our case as it abides by the protocols and regulations. We can consider this scenario as a part of a larger scenario in which ST sends both confidential and common messages to both SRs at different time intervals in the long-run. We further assume that the PUs are not aware of the coding strategies of the SUs.

In the above model, the generated data sequences are initially stored in the data buffer, and then transmitted by ST in frames of $T$ seconds over the wireless channel. During the data transmission, the channel input-output relations at the $j^{th}$ time instant are
\begin{alignat}{4}
y_{m}(j)&=h_{m}(j)x(j)&+&n_{m}(j)&+&s_{m}(j),\label{channel_state_1_busy}\\
y_{e}(j)&=h_{e}(j)x(j)&+&n_{e}(j)&+&s_{e}(j),\label{channel_state_2_busy}
\intertext{if the PUs are active, and}
y_{m}(j)&=h_{m}(j)x(j)&+&n_{m}(j),&\text{ }&\label{channel_state_1_idle}\\
y_{e}(j)&=h_{e}(j)x(j)&+&n_{e}(j),&\text{ }&\quad\text{for }
j=1,2,...,\label{channel_state_2_idle}
\end{alignat}
if the PUs are absent. Above, $x(j)$ denotes the complex-valued channel input, and $y_{m}(j)$ is the complex-valued channel output at $\sr_{m}$, while $y_{e}(j)$ is the complex-valued channel output at $\sr_{e}$. In (\ref{channel_state_1_busy}) and (\ref{channel_state_1_idle}), $h_{m}(j)$ represents the fading coefficient between ST and $\sr_{m}$, and in (\ref{channel_state_2_busy}) and (\ref{channel_state_2_idle}), $h_{e}(j)$ represents the fading coefficient between ST and $\sr_{e}$. They both are assumed to be zero-mean, circularly symmetric, complex Gaussian distributed random
variables with variances $\mathbb{E}\{|h_{m}(j)|^2\}=\mathbb{E}\{z_{m}(j)\}=\sigma_{m}^2$ and $\mathbb{E}\{|h_{e}(j)|^2\}=\mathbb{E}\{z_{e}(j)\}=\sigma_{e}^2$. Note that $z_{m}(j)$ and $z_{e}(j)$ are the magnitude squares of instantaneous fading coefficients. In the above equations, $\{n_{m}(j)\}$ and $\{n_{e}(j)\}$ are sequences of additive thermal random noise samples at $\sr_{m}$ and $\sr_{e}$, that are also zero-mean, circularly symmetric, complex Gaussian distributed with variances $\mathbb{E}\{|n_{m}(j)|^2\}=\sigma_{nm}^2$ and $\mathbb{E}\{|n_{e}(j)|^2\}=\sigma_{ne}^2$, respectively. In (\ref{channel_state_1_busy}) and (\ref{channel_state_2_busy}), $s_{m}(j)$ arriving at $\sr_m$ and $s_e(j)$ arriving at $\sr_{e}$ denote the active PUs' faded signals. We show the average power levels of $s_{m}(j)$ and $s_{e}(j)$ with $\sigma_{sm}^2$ and $\sigma_{se}^2$, respectively.

We further consider a block-fading channel model, and assume that the fading coefficients stay constant for a frame duration of $T$ seconds, and change independently from one frame to another. Besides, we also assume that the activities of the PUs stay the same in each frame, and that they have likewise an independent activity state change from one frame to another. We remark that the PUs exist with probability $\rho$ in each frame. Meanwhile, ST, $\sr_{m}$, and $\sr_{e}$ experience the interference caused by the PUs contemporaneously when the PUs are active, but not at the same average power level. We finally express that the available bandwidth is $B$, and so is the symbol rate assumed to be $B$ complex symbols per second.

\subsection{Interference Power Constraints}
Initially, the SUs perform channel sensing to verify the activities of the PUs, and then, depending on the channel sensing results, they choose their transmission power policies. In more detail, if the channel is sensed to be busy, ST sends data with an instantaneous power policy of $P_{b}(j)$, whereas, if the channel is sensed as idle, it sends with an instantaneous power policy of $P_{i}(j)$. Additionally, we consider a practical scenario in which channel sensing errors such as miss-detections and false alarms possibly occur. Hence, we denote the correct-detection probability by $P_{d}$ which is the probability of declaring the channel as busy while the channel is actually busy, and the false alarm probability by $P_{f}$ which is the probability of declaring the channel as busy while the channel is actually idle. Here, the busy state indicates that the PUs are active, and the idle state indicates that there is not any active PU in the transmission environment. We further notice that as a result of channel sensing errors, ST deploys policies $P_{b}(j)$ and $P_{i}(j)$ with probabilities $P_{d}$ and $(1-P_{d})$, respectively, given that the PUs are actually active. Therefore, the power interference constraint is imposed on not only $P_{b}(j)$ but also $P_{i}(j)$. Thus, the combined interference power constraint is given as
\begin{equation}\label{interference_power_constraint}
P_{d}\mathbb{E}\{P_{b}(j)\}+(1-P_{d})\mathbb{E}\{P_{i}(j)\}\leqslant P_{int}
\end{equation}
where the expectation $\mathbb{E}\{\cdot\}$ is taken with respect to the fading coefficients $z_{m}$ and $z_{e}$, and $P_{int}$ is the average power interference constraint\footnote{$P_{int}$ is considered as the average power interference constraint normalized over average fading power and path loss of the channel between ST and the PRs.}. Now, let $\mu_{b}(j)=\frac{P_{b}(j)}{P_{int}}$ and $\mu_{i}(j)=\frac{P_{i}(j)}{P_{int}}$ be the transmission power policies normalized over $P_{int}$. Hence, we can rewrite (\ref{interference_power_constraint}) as follows:
\begin{equation}\label{interference_power_constraint_mu}
P_{d}\mathbb{E}\{\mu_{b}(j)\}+(1-P_{d})\mathbb{E}\{\mu_{i}(j)\}\leqslant1.
\end{equation}
Since, the average transmission power of ST is limited by $P_{int}$, we define the following signal-to-noise ratio at $\sr_{m}$ as $\tSNR=\frac{P_{int}}{B\sigma_{nm}^2}$.
\section{Secrecy Rates}\label{secrecy_rates_section}
Initially, we consider four different transmission scenarios regarding the channel sensing decision and its correctness:
\begin{enumerate}
  \item Channel is busy, detected as busy (correct detection),
  \item Channel is busy, detected as idle (miss-detection),
  \item Channel is idle, detected as busy (false alarm),
  \item Channel is idle, detected as idle (correct detection).
\end{enumerate}
We can easily see that ST will send data with the power policy $\mu_{b}(j)$ in Scenarios 1 and 3,
and $\mu_{i}(j)$ in Scenarios 2 and 4. Hence, assuming the interference caused by the PUs, $s_m(j)$ and $s_e(j)$, as additional Gaussian noise, the instantaneous secure channel capacities in the above four scenarios are expressed as
\begin{align}
C_{k}(j)=\big[&\underbrace{B\log_{2}\left(1+\zeta_{m,k}(j)\right)}_{C_{mk}(j)}-\underbrace{B\log_{2}\left(1+\zeta_{e,k}(j)\right)}_{C_{ek}(j)}\big]^+\nonumber\\&\text{ bits/s for }k=\{1,2,3,4\}, \label{capacity_ifadeleri}
\end{align}
where
\begin{align*}
&\zeta_{m,1}(j)=\frac{z_{m}(j)\tSNR\mu_{b}(j)}{\beta},&\zeta_{e,1}(j)=\frac{z_{e}(j)\alpha_b\tSNR\mu_{b}(j)}{\beta},\\
&\zeta_{m,2}(j)=\frac{z_{m}(j)\tSNR\mu_{i}(j)}{\beta},&\zeta_{e,2}(j)=\frac{z_{e}(j)\alpha_b\tSNR\mu_{i}(j)}{\beta},\\
&\zeta_{m,3}(j)=z_{m}(j)\tSNR\mu_{b}(j),&\zeta_{e,3}(j)=z_{e}(j)\alpha_{i}\tSNR\mu_{b}(j),\\
&\zeta_{m,4}(j)=z_{m}(j)\tSNR\mu_{i}(j),&\zeta_{e,4}(j)=z_{e}(j)\alpha_{i}\tSNR\mu_{i}(j).
\end{align*}
Above, $\alpha_{b}=\frac{\sigma_{nm}^2+\sigma_{sm}^2}{\sigma_{ne}^2+\sigma_{se}^2}$, $\beta=1+\frac{\sigma_{sm}^2}{\sigma_{nm}^2}$, $\alpha_{i}=\frac{\sigma_{nm}^2}{\sigma_{ne}^2}$, and $[a]^+=\max(0,a)$. We observe that the first expression in the right-hand-side of the above equation, $C_{mk}(j)$, is in fact the instantaneous channel capacity between ST and $\sr_{m}$, and the second expression, $C_{ek}(j)$, is the instantaneous channel capacity between ST and $\sr_{e}$. It is clearly seen that when $C_{ek}(j)$ is greater than or equal to $C_{mk}(j)$, the instantaneous secure channel capacity will be zero. Therefore, when the channel of $\sr_{e}$ is stronger than the channel of $\sr_{m}$, ST is not required to transmit any secret data to $\sr_m$ (i.e., when $z_{e}(j)\alpha_{b}\geq z_{m}(j)$, $\mu_{b}(j)=0$ in Scenario 1 and $\mu_{i}(j)=0$ in Scenario 2, and when $z_{e}(j)\alpha_{i}\geq z_{m}(j)$, $\mu_{b}(j)=0$ in Scenario 3 and $\mu_{i}(j)=0$ in Scenario 4.).

Now, let $r_b(j)$ and $r_i(j)$ denote the secure transmission rates, and let $r_{be}(j)$ and $r_{ie}(j)$ denote the confusion rates, when the channel is sensed as busy and idle, respectively. Hence, when the channel of $\sr_m$ is stronger than the channel of $\sr_e$, we assume that with a secure transmission rate lower than or equal to the instantaneous secure channel capacity (i.e., $r_{b}(j)\leq C_{1}(j)$ and $r_{b}(j)\leq C_{3}(j)$ in Scenarios 1 and 3, respectively, and $r_{i}(j)\leq C_{2}(j)$ and $r_{i}(j)\leq C_{4}(j)$ in Scenarios 2 and 4, respectively), and a confusion rate greater than or equal to the instantaneous channel capacity between ST and $\sr_e$ (i.e., $r_{be}(j)\geq C_{1e}(j)$ and $r_{be}(j)\geq C_{3e}(j)$ in Scenarios 1 and 3, respectively, and $r_{ie}(j)\geq C_{2e}(j)$ and $r_{ie}(j)\geq C_{4e}(j)$ in Scenarios 2 and 4, respectively), ST will be able to reliably transmit data to $\sr_{m}$ without revealing any information to $\sr_{e}$ as long as the sum of secure transmission and confusion rates is smaller than the channel capacity between ST and $\sr_m$ (i.e., $C_{1m}(j)\geq r_{bm}(j)=r_{b}(j)+r_{be}(j)$ and $C_{3m}(j)\geq r_{bm}(j)$ in Scenarios 1 and 3, respectively, and $C_{2m}(j)\geq r_{im}(j)=r_{i}(j)+r_{ie}(j)$ and $C_{4m}(j)\geq r_{im}(j)$ in Scenarios 2 and 4, respectively). In this case, $\sr_m$ will be able to decode both the confusion and secure data, but $\sr_e$ will be blinded by the confusion data and unable to decode the secure data.

Since the SUs rely on the channel sensing results, and ST knows both channel fading coefficients, and the SRs know their corresponding channel fading coefficients, one of the best transmission strategies could be that when the PUs are detected as active, the instantaneous secure transmission rate in one frame is
\begin{align*}
&r_{b}(j)=C_{1}(j)\\&=\big[\underbrace{B\log_{2}\left(1+\zeta_{m,1}(j)\right)}_{r_{bm}(j)=C_{m1}(j)}
-\underbrace{B\log_{2}\left(1+\zeta_{e,1}(j)\right)}_{r_{be}(j)=C_{e1}(j)}\big]^+\text{ bits/s},
\end{align*}
and that when the PUs are detected as idle, it is
\begin{align*}
&r_{i}(j)=C_{4}(j)\\&=\big[\underbrace{B\log_{2}\left(1+\zeta_{m,4}(j)\right)}_{r_{im}(j)=C_{m4}(j)}
-\underbrace{B\log_{2}\left(1+\zeta_{e,4}(j)\right)}_{r_{ie}(j)=C_{e4}(j)}\big]^+\text{ bits/s}.
\end{align*}
Therefore, we have the instantaneous secure transmission rates $r_{b}(j)$ equal to $C_{1}(j)$ in Scenarios 1 and 3, and $r_{i}(j)$ equal to $C_{4}(j)$ in Scenarios 2 and 4. It is clearly seen that in Scenarios 1 and 4, since the sum of secure transmission rate and confusion rate is equal to the instantaneous channel capacity between ST and $\sr_m$ (i.e., $r_{bm}(j)=r_{b}(j)+r_{be}(j)=C_{m1}(j)$ and $r_{im}(j)=r_{i}(j)+r_{ie}(j)=C_{m4}(j)$), and since the instantaneous confusion rate is equal to the instantaneous channel capacity between ST and $\sr_e$ (i.e., $r_{be}(j)=C_{e1}(j)$ and $r_{ie}(j)=C_{e4}(j)$), there will be no transmission and security outages, and a reliable and secure communication will be provided. As a result, the effective secure transmission rate in one frame is $Tr_{b}(j)$ in Scenario 1, and $Tr_{i}(j)$ in Scenario 4. On the other hand, in Scenarios 2 and 3, we do have different situations. For instance, 
$r_{im}(j)>C_{m2}(j)$ in Scenario 2. Therefore, no reliable communication will be performed because $\sr_{m}$ will not be able to decode the confusion and transmitted data, and as a result the effective secure transmission rate is zero. It is, thence, assumed that a simple automatic repeat request mechanism is invoked in order to ensure that erroneous data is retransmitted. In addition, we observe that $r_{ie}(j)>C_{e2}(j)$. 
Correspondingly, $\sr_{e}$ will not be able to decode the transmitted data intended for $\sr_{m}$ too. Therefore, there is no security outage in this scenario as well. However, we see that $r_{bm}(j)<C_{m3}(j)$ 
in Scenario 3, which means that $\sr_{m}$ will be able to decode the confusion and transmitted data. At the same time, we also mark that $r_{be}(j)<C_{e3}(j)$. 
Particularly, the confusion created by ST has a smaller rate than the required value, $C_{e3}(j)$. Therefore, it is also likely that $\sr_{e}$ can decode the transmitted data especially when $z_{e}\beta\alpha_{i}\geq z_{m}$ (i.e., $C_{e3}(j)\geq r_{bm}(j)$), which will jeopardize the secrecy of the transmitted data. Therefore, in order to guarantee the information-theoretic security, we propose the following transmission rate policy when PUs are detected as active:
\begin{align*}
r_{b}(j)=\big[\underbrace{B\log_{2}\left(1+\zeta_{m,1}(j)\right)}_{r_{bm}(j)=C_{m1}(j)}-&\underbrace{B\log_{2}\left(1+\zeta_{e,3}(j)\right)}_{r_{be}(j)=C_{e3}(j)}\big]^+\text{ bits/s}.
\end{align*}

Now, when the channel is detected as busy while $z_{m}(j)>z_{e}(j)\beta\alpha_{i}$, the security will always be provided in Scenario 1, because $\sr_{e}$ will not be able to decode the transmitted data since $r_{be}(j)>C_{e1}(j)$. Meanwhile, the security in Scenario 3 will also be provided, because $r_{be}(j)=C_{e3}(j)$. Note that $\sr_{m}$ will securely decode its data since $C_{m1}(j)=r_{bm}(j)$ and $C_{m3}(j)>r_{bm}(j)$ in Scenarios 1 and 3, respectively.

\section{Effective Secure Capacity}\label{effective_secure_rate_section}
In this section, having characterized the instantaneous secure transmission rates, we investigate the secure throughput for the aforementioned fading cognitive radio wiretap channel under statistical QoS constraints. In \cite{wu_negi}, the authors defined effective capacity as the maximum constant arrival rate that a given service process can support in order to guarantee a desired statistical QoS specified with the QoS exponent $\theta$. Defining $Q$ as the stationary queue length and $\theta$ as the decay rate of the tail distribution of the queue length $Q$, we can express the following:
\begin{equation}
\lim_{q\to\infty}\frac{\log{\Pr(Q\geq q)}}{q} = -\theta.
\end{equation}
Therefore, we have the following approximation for larger $q$: $\Pr(Q\geq q)\approx e^{-\theta q}$, which means that larger $\theta$ refers to strict buffer constraints, and smaller $\theta$ implies looser constraints. Furthermore, it is shown in \cite{liu_chamberland} that $\Pr(D\geq d)\leq c\sqrt{\Pr(Q\geq q)}$ for constant arrival rates, where $D$ denotes the steady-state delay experienced in the buffer, and $c$ is a positive constant. In the above formulation, $q=ad$ where $a$ is the source arrival rate. Therefore, effective capacity can provide us the maximum arrival rate when the system is subject to the statistical queue length or delay constraints in the forms of $\Pr(Q\geq q)\leq e^{-\theta q}$ or $\Pr(D\geq d)\leq ce^{-\theta ad/2}$, respectively.

The effective capacity for a given QoS exponent $\theta$ is given by
\begin{equation}
-\lim_{t\to\infty}\frac{1}{\theta t}\log_{e}E\{e^{-\theta
S(t)}\}=\frac{\Lambda(-\theta)}{\theta}
\end{equation}
where $S(t)=\sum_{l=1}^{t}r(l)$ is the time-accumulated service process, and $r(l)$ for $l=1,2,...$ is the discrete-time, stationary and ergodic stochastic service process. We note that $\Lambda(\theta)$ is the asymptotic log-moment generating function of $S(t)$, and is given by
\begin{equation}
\Lambda(\theta)=\lim_{t\to\infty}\frac{1}{t}\log E\left\{e^{\theta
S(t)}\right\}.
\end{equation}
\begin{figure}
\begin{center}
\includegraphics[width=0.35\textwidth]{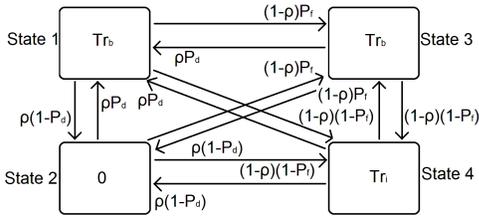}
\caption{State transition diagram.}\label{res:res_4}
\end{center}
\end{figure}

Under block-fading assumption, the securely transmitted number of bits is $Tr_{b}(j)$ or $Tr_{i}(j)$ in the $l^{th}$ transmission frame, if the transmission policy $\mu_{b}(j)$ or $\mu_{i}(j)$ is chosen, respectively, and if the data is transmitted successfully. In the sequel, since the transmission rate is constant during one frame, and the channel fading coefficients and the activities of the PUs change independently from one frame to another, we will drop the time index $j$. In the above model, when the channel is detected as busy in Scenarios 1 and 3, the service rate is $Tr_{b}$, whereas, when the channel is detected as idle in Scenario 4, the service rate is $Tr_{i}$ in one frame. In Scenario 2, the service rate is zero.

In Figure \ref{res:res_4}, we display the state transition diagram of the cognitive radio wiretap channel. On the top-left (Scenario 1) and top-right (Scenario 3), we see the states in which ST sends data reliably with a secure transmission rate $r_{b}$ in one frame, whereas on the lower-right (Scenario 4), we have the state in which data is transmitted with a secure transmission rate $r_{i}$. On the other hand, on the lower-left (Scenario 2), due to miss-detection when the PUs are active, the effective secure transmission rate is 0. Moreover, state transition probabilities are also shown in the figure. For instance, state transition from any state to State 1 is $\rho P_{d}$. Similarly, state transition probabilities to States 2, 3, and 4 are $\rho(1-P_{d})$, $(1-\rho)P_{f}$, and $(1-\rho)(1-P_{f})$, respectively. This is due to the fact that the activities of the PUs and the channel fading coefficients change independently from one frame to another.

Following the steps in \cite{sami_1}, we can express the effective secure capacity, which is the maximum constant arrival rate of secret data arriving at ST, as
\begin{align}
R_{e}(\theta)=\frac{-1}{\theta
BT}\log_{e}\mathbb{E}\left[p_{b}e^{-\theta Tr_{b}}
+p_{i}e^{-\theta Tr_{i}}+p_{0}\right]\text{ bits/s/Hz.}\label{effective_capacity}
\end{align}
where $p_{b}=\rho P_{d}+(1-\rho)P_{f}$, $p_{i}=(1-\rho)(1-P_{f})$, and
$p_{0}=\rho(1-P_{d})$.

\vspace{-0.15cm}
\section{Optimal Transmission Power Control}\label{optimal_transmission_power_section}
After characterizing the effective secure capacity, we turn our attention to optimal transmission power policies that will maximize the expression in (\ref{effective_capacity}). The following result provides the optimal power allocation policies.

\begin{theo}\label{optimal_power_allocation_policies}
The optimal power allocation policies, $\mu_b$ and $\mu_i$, that maximize the effective secure capacity are given by
\begingroup
\allowdisplaybreaks
\begin{align}
\mu_b&=\left\{
\begin{array}{ccc}
X:X=H_b(X),&z_{m}-z_{e}\beta\alpha_{i}>\frac{\gamma_{0}P_{d}\beta}{p_{b}},\\
0,&\hbox{otherwise,}
\end{array}\right.\label{mu_b_optimal_ori}\\
\intertext{and}
\mu_i&=\left\{
\begin{array}{ccc}
X:X=H_i(X),&z_{m}-z_{e}\alpha_{i}>\frac{\gamma_0(1-P_{d})}{p_{i}},\\
 0,&\quad\hbox{otherwise,}
\end{array}
\right.\label{mu_i_optimal_ori}
\end{align}
\endgroup
respectively, where
\begingroup
\allowdisplaybreaks
\begin{align}
H_b(X)&=\frac{(z_{m}-z_{e}\beta\alpha_{i})\sqrt{1+\Phi}-(z_m+z_e\beta\alpha_i)}{2z_mz_e\alpha_i\tSNR},\label{mu_b_optimal}\\
H_i(X)&=\frac{(z_m-z_e\alpha_i)\sqrt{1+\Psi}-(z_m+z_e\alpha_i)}{2z_mz_e\alpha_i\tSNR},\label{mu_i_optimal}
\end{align}
\endgroup
$\Phi=\frac{4z_mz_e\alpha_ip_bf(X)}{\gamma_0P_d(z_m-z_e\beta\alpha_i)}$, $\Psi=\frac{4z_mz_e\alpha_{i}p_ig(X)}{\gamma_0(1-P_d)(z_m-z_e\alpha_i)}$, $f(X)=\left(\frac{\beta+z_{m}\tSNR X}{\beta +z_{e}\beta\alpha_{i}\tSNR X}\right)^{-\kappa}$, $g(X)=\left(\frac{1+z_{m}\tSNR X}{1+z_{e}\alpha_{i}\tSNR X}\right)^{-\kappa}$, and $\kappa=\frac{\theta TB}{\log2}$. $\gamma_0$ is the power threshold value in the power adaptation policies, and it can be obtained from the average interference power constraint given in (\ref{interference_power_constraint_mu}) through numerical techniques\footnote{We note that when $z_{e}=0$, the transmitter does not need to consider the above policies since there is no direct link to $\sr_e$. In such a case, security is guaranteed at any value of $z_m$.}.
\end{theo}

Now, we can easily find out that (\ref{mu_b_optimal}) and (\ref{mu_i_optimal}) are monotonically decreasing convex functions of $X$ for $X\geq0$, 
and $H_{b}(0)$ and $H_{i}(0)$ are greater than 0. Therefore, we provide the following simple algorithm to obtain optimal power policies:
\begingroup
\captionof{algorithm}{Power Control\label{algo:power}}
\begin{algorithmic}[1]
\State Initialize $\gamma_0$;
\If{$z_m-z_e\beta\alpha_i\leq\frac{\gamma_0P_d\beta}{p_b}$}\label{xsdsad}
\State $\mu_b=0$;
\Else
\State Initialize $\mu_b^{0}=0$, $\mu_b^{2}=H_b(0)$, $\mu_b^{1}=(\mu_b^{0}+\mu_b^{2})/2$;
\While{Until $\mu_b$ converges}
\State $\mu_b=H_b(\mu_b^{1})$;
\If{$\mu_b>\mu_b^{2}$}\label{buradan}
\State $\mu_b^{0}=\mu_b^{1}$, $\mu_b^{1}=(\mu_b^{0}+\mu_b^{2})/2$;
\ElsIf{$\mu_b>\mu_b^{1}$}
\State $\mu_b^{0}=\mu_b^{1}$, $\mu_b^1=\mu_b^{2}=\mu_b$;
\ElsIf{$\mu_b>\mu_b^{0}$}
\State $\mu_b^{2}=\mu_b^{1}$, $\mu_b^0=\mu_b^{1}=\mu_b$;
\Else
\State $\mu_b^{2}=\mu_b^{1}$, $\mu_b^{1}=(\mu_b^{0}+\mu_b^{2})/2$;
\EndIf\label{buraya}
\EndWhile
\EndIf
\If{$z_m-z_e\alpha_i\leq\frac{\gamma_0(1-P_d)}{p_i}$}
\State $\mu_i=0$;
\Else
\State Initialize $\mu_i^{0}=0$, $\mu_i^{2}=H_i(0)$, $\mu_i^{1}=(\mu_i^{0}+\mu_i^{2})/2$;
\While{Until $\mu_i$ converges}
\State $\mu_i=H_i(\mu_i^{1})$;
\State ... Follow Steps \ref{buradan}-\ref{buraya} for $\mu_i$ ...;
\EndWhile
\EndIf
\If{(\ref{interference_power_constraint_mu}) is satisfied}
\State Declare $\mu_b$ and $\mu_i$;
\Else
\State Update $\gamma_0$ and return to Step \ref{xsdsad};
\EndIf
\end{algorithmic}
\endgroup

\section{Numerical Results}\label{numerical_results_section}
In this section, we present the numerical results. Unless indicated otherwise, we consider the following parameter values. The channel is assumed to be busy with probability $\rho=0.1$. The noise variances are assumed to be $\sigma_{nm}^{2}=1$ and $\sigma_{ne}^2=1$, and the average power of the PUs' signals arriving at the SRs are $\sigma_{sm}^2=1$ and $\sigma_{se}^2=1$. We further assume that the fading coefficients have unit variances $\sigma_m^2=\sigma_e^2=1$ in Rayleigh fading environment. 
Channel bandwidth is $B=100$ Hz, and the frame duration is $T=1$ second. 

\begin{figure}
\begin{center}
\includegraphics[width=\figsize\textwidth]{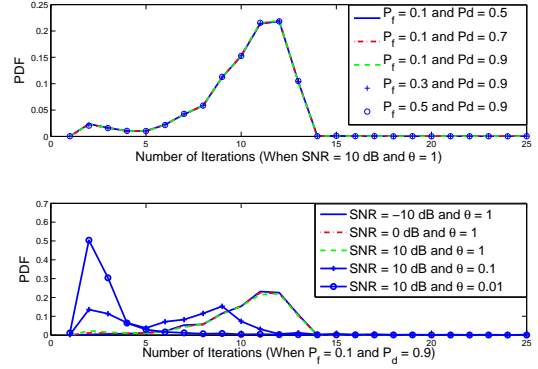}
\caption{Probability distribution of the number of iterations required for Algorithm \ref{algo:power} to converge to the solution.}\label{fig:fig_1}
\end{center}
\end{figure}

In Figure \ref{fig:fig_1}, we demonstrate the performance of Algorithm \ref{algo:power} under different channel conditions and QoS requirements. In the upper part of the figure, we display the probability distribution of number of iterations required for convergence for different values of $P_f$ and $P_d$ when $\tSNR=10$ dB and QoS exponent $\theta=1$. We notice that channel sensing performance do not affect the required number of iterations. The required number of iterations is often less than 14. On the other hand, in the lower part of the figure, we observe that while the value of $\theta$ has an impact on the speed of the algorithm, $\tSNR$ does not have any effect. With decreasing $\theta$ (less strict QoS requirements), the number of iterations also decreases. For instance, when $\theta=0.01$, mostly with less than 5 iterations, the algorithm converges to the solution.

\begin{figure}
\begin{center}
\includegraphics[width=\figsize\textwidth]{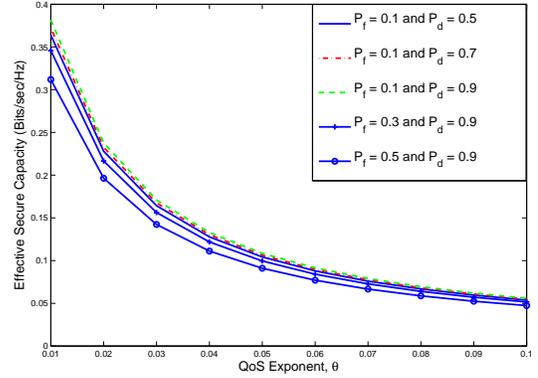}
\caption{Maximized effective secure capacity vs. QoS exponent $\theta$ for different values of $P_f$ and $P_d$.}\label{fig:fig_2}
\end{center}
\end{figure}
\begin{figure}
\begin{center}
\includegraphics[width=\figsize\textwidth]{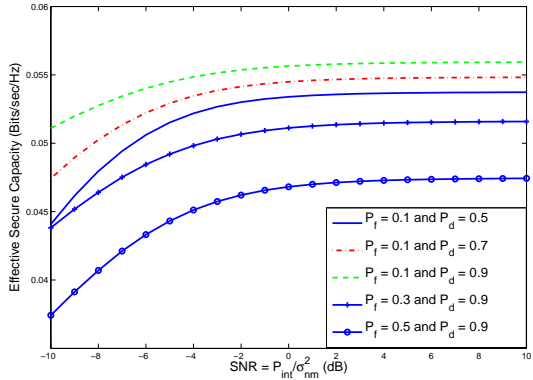}
\caption{Maximized effective secure capacity vs. $\tSNR$ for different values of $P_f$ and $P_d$ when $\theta=0.1$.}\label{fig:fig_3}
\end{center}
\end{figure}

In Figure \ref{fig:fig_2}, we plot the effective secure capacity as a function of $\theta$ for different $P_{f}$ and $P_{d}$ values when $\tSNR=10$ dB. As expected, effective secure capacity increases with decreasing $\theta$, and better channel sensing performance enhances the transmission throughput, e.g., $P_f=0.1$ and $P_d=0.9$. We interestingly note that the gains attained through boosting channel sensing performance tend to diminish with increasing $\theta$. In Figure \ref{fig:fig_3}, we plot the effective secure capacity as a function of $\tSNR$ for different channel sensing results when $\theta=0.1$. We point out that effective secure capacity does not escalate with the looser interference power constraints after certain $\tSNR$($P_{int}$) value. In contrast to results in Fig. \ref{fig:fig_2}, we see that channel sensing performance has a considerable effect in transmission throughput.

\begin{figure}
\begin{center}
\includegraphics[width=\figsize\textwidth]{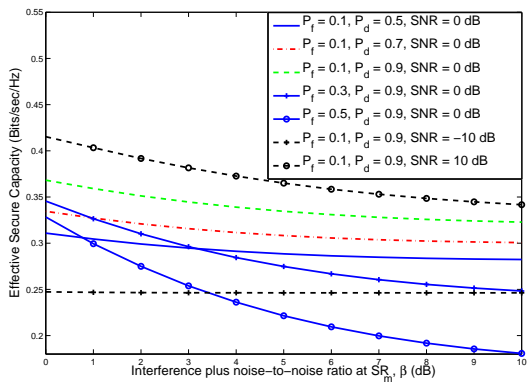}
\caption{Maximized effective secure capacity vs. $\beta$ for different $P_{f}$, $P_d$, and $\tSNR$ values when $\theta=0.01$.}\label{fig:fig_4}
\end{center}
\end{figure}
We recall that $\beta=1+\frac{\sigma_{sm}^2}{\sigma_{nm}^2}$ is the interference plus noise-to-noise ratio at $\sr_m$ which can be considered as an important parameter that will help us analyze the effects of hidden-terminal problem\footnote{Hidden-terminal or hidden-node problem occurs when PUs' signals are seen from a SU but not other SUs that communicate with that SU.} in cognitive radio wiretap channels. We know that the minimum value of $\beta$ is 1. In Figure \ref{fig:fig_4}, we plot the effective secure capacity vs. $\beta$. We observe that channel sensing results have a crucial role in the transmission throughput. Although the PUs have strong signal power, they may not be detected due to hidden-terminal effects. Therefore, $P_d$ might be very low. In order to increase $P_{d}$, the SUs may be required to increase the false alarm probability, $P_{f}$. As a result, even with high $\tSNR$, when $\beta$ has a considerable value the transmission throughput may be lower than the throughput obtained with low $\tSNR$ while having better channel sensing results. This is due to increased $P_f$. For instance, when $P_f=0.5$, $P_d=0.9$ and $\tSNR=0$ dB, the effective secure capacity is lower than the effective secure capacity obtained when $P_f=0.1$, $P_d=0.9$ and $\tSNR=-10$ when $\beta$ is high.

\section{Conclusion}\label{conclusion_section}
In this paper, we have analyzed the effective secure capacity in cognitive radio wiretap channels in order to investigate interactions among information-theoretic security, channel sensing performance and QoS requirements. We have initially constructed channel scenarios, and have provided instantaneous secure channel capacities in each scenario. We have established transmission strategies regarding channel sensing results and channel fading coefficients. We have acquired optimal transmission power policies, and have provided the algorithm that ascertains these policy values as a function of channel fading coefficients, channel sensing results, interference power constraints and QoS requirements. Finally, we have presented numerical results, and have highlighted the consequences of hidden-terminal on effective secure capacity.

\end{document}